\documentclass[epj]{svjour}

\usepackage{graphics}

\newcommand{\mtbox}[1]{{\mbox{\scriptsize #1}}}

\begin{document}

\title{Chromofields of Strings and Baryons 
  \thanks{Work supported by BMBF}
  }
\author{Gunnar Martens \and Carsten Greiner \and Stefan
  Leupold \and Ulrich Mosel%
  }                     %
\institute{Inst.~f.~Theo.~Physik, Universit\"at Giessen,
  Heinrich--Buff--Ring 16, 35392 Giessen   
}

\date{Received: date / Revised version: date}

\abstract{
We calculate color electric fields of quark/antiquark ($\bar{q}q$) and 3
quark ($qqq$) systems within the chromodielectric model (CDM). We
explicitly evaluate the string tension of flux tubes in the
$\bar{q}q$--system and analyze their profile. To reproduce results
of lattice calculations we use a bag pressure $B = (320\mbox{MeV})^4$
from which an effective strong coupling constant $\alpha_{s} \approx
0.3$ follows. With these parameters we get a  $Y$ shaped configuration
for large $qqq$--systems.  
}

\maketitle

\section{Introduction}
\label{sec:intro}
Quantum chromodynamics (QCD) is the widely ac\-cep\-ted theory for the
dynamics of quarks and gluons. Despite its success in the regime of
high momemtum transfer it remains an outstanding task to explain the
low energy behavior of hadrons within QCD. Only in the last 10 years
lattice QCD (lQCD) has found detailed evidence for the  confinement of
quarks 
in hadrons 
\cite{Bali:2000gf} but it still fails to give a dynamical description
of this phenomenon. It is therefore necessary to rely on 
models, capable to describe confinement dynamically on the one hand
and to reproduce static results of lQCD on the other hand. 

In this talk we present static calculations within the Chromodielectric
Model \cite{Friedberg:1977eg,Friedberg:1977xf,Traxler:1998bk}, namely
the detailed analysis of quark--antiquark strings and three--quark
configurations. 

\section{Phenomenology of the Model}
\label{sec:phenomenology-model}
\begin{figure}
  \begin{center}
    \resizebox{0.24\textwidth}{!}{%
      \includegraphics{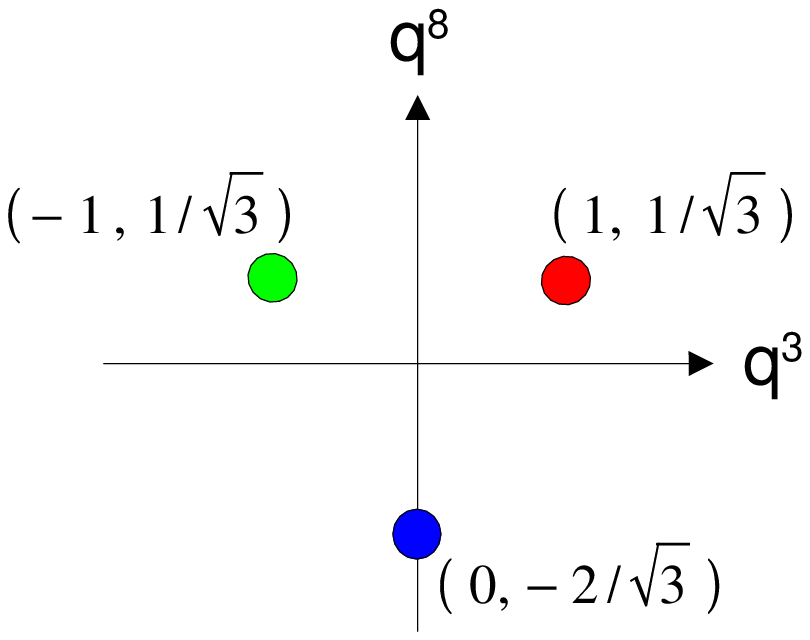}
      }
    \hfill
    \resizebox{0.24\textwidth}{!}{%
      \includegraphics{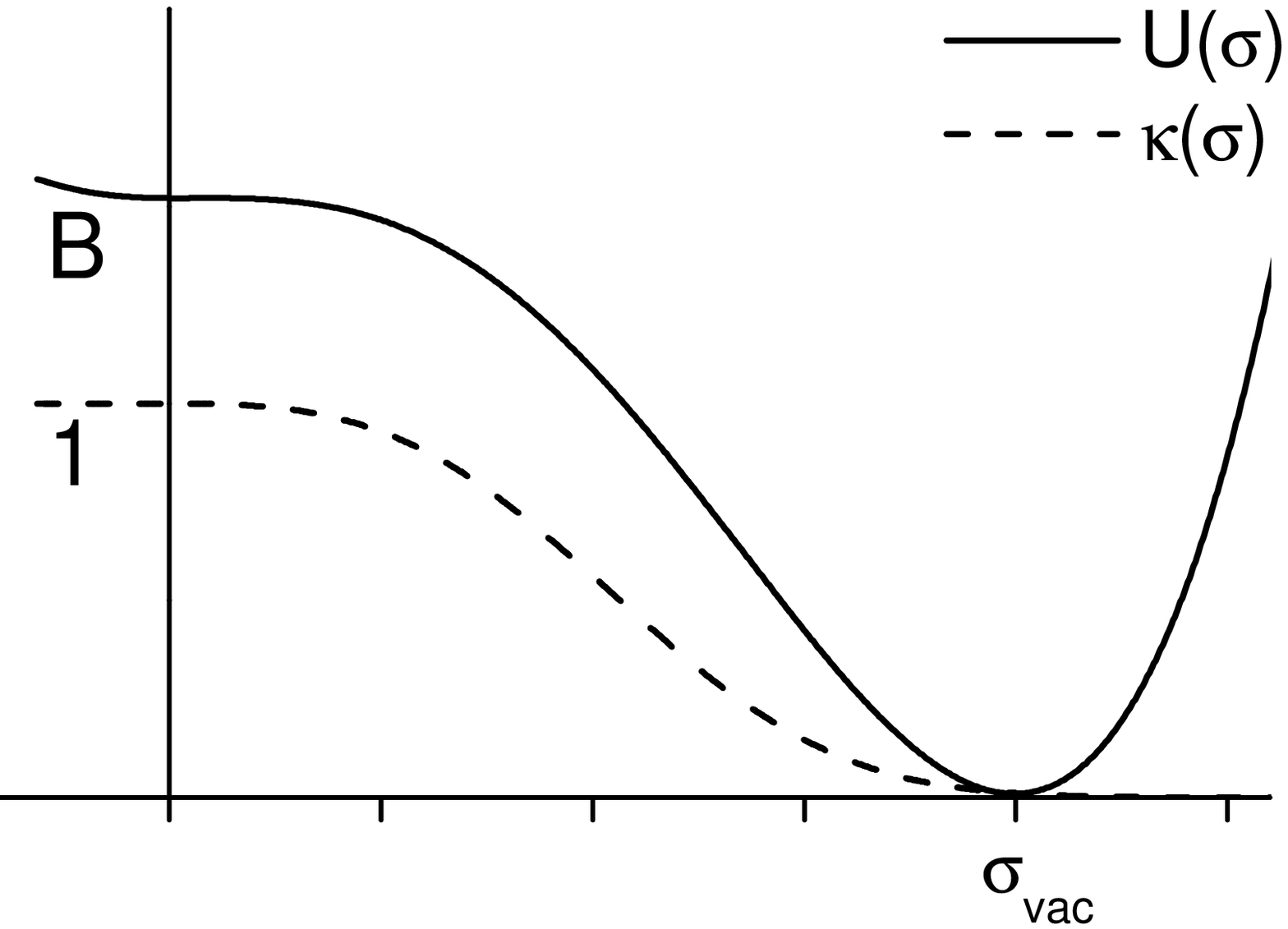}
      }
    \caption{Left: The color charges of the green (left), red (right)
      and blue (lower) quark. Right: The dielectric function
      $\kappa(\sigma)$ (dashed) and the scalar potential
      $U(\sigma)$. 
      }
    \label{fig:ukappa}       
  \end{center}
\end{figure}

In the Chromodielectric Model (CDM) it is assumed, that
the vacuum of QCD behaves in the long range limit as a perfect color
dielectric medium with vanishing dielectric constant $\kappa = 0$.
The medium is generated through the non-abelian part of the
gluonic sector of QCD which is represented in CDM as a scalar color
singlet field $\sigma$. The remaining two abelian gluon fields
are able to propagate through this medium. The scalar field $\sigma$
is driven by a scalar potential $U(\sigma)$ (see
fig.~\ref{fig:ukappa}) 
which exhibits
two (quasi) stable points, separating the non-perturbative, perfect
dielectric phase where  $\sigma = \sigma_{\mtbox{vac}}$, from the
perturbative phase with $\kappa = 1$, where the color fields can
propagate freely and $\sigma = 0$.

In our description quarks are treated classically and the gluons
are coupled to the quark current $j^{\mu,a} $. This results in the
following Lagrangian 
\begin{eqnarray}
  \label{eq:lagrangian}
  {\cal L} &=& {\cal L}_q + {\cal L}_g + {\cal L}_{\sigma}\\
  {\cal L}_q &=& - \sum\limits_k 
  m_k \sqrt{1 - \dot{\vec{x}}_k^2} \;
  w\left(\vec{x} - \vec{x}_k(t)\right) \\
  && -g_s \;j_\mu^a \,A^{\mu,a}\nonumber\\
  {\cal L}_g &=& - {\textstyle \frac{1}{4}} {\kappa(\sigma)}
  F_{\mu\nu}^{a} F^{\mu\nu, {a}}\\
  {\cal L}_{\sigma} &=& {{\textstyle \frac{1}{2}}
    \partial_\mu\sigma\partial^\mu\sigma  - U(\sigma) }\\
  F^{\mu\nu, { a}} &=& \partial^\mu A^{\nu,  a} 
  - \partial^\nu A^{\mu,  a}, \;\;\; a \in\{3,8\}\\
  j^{\mu,a} &=& \sum\limits_k q_k^a\; u_k^\mu\; 
  w\left(\vec{x} - \vec{x}_k(t)\right) = (\rho^a,
  \vec{\jmath}^{\;a})
\end{eqnarray}
with $u_k^\mu$ being the 4-velocity of particle $k$ with classical charge
$q_k^a$ (see fig.~\ref{fig:ukappa}) and extension $w\left(\vec{x} -
  \vec{x}_k(t)\right)$.
The scalar potential $U(\sigma)$ is
chosen to be of a quartic form and is shown in
fig.~\ref{fig:ukappa}. In this work $U(\sigma)$ has no relative
maximum between $\sigma=\sigma_{\mtbox{vac}}$ and $\sigma = 0$ and $U$
is determined through the bag pressure $B = U(0)$ and
$\sigma_{\mtbox{vac}}$ alone. The dielectric function is of the form
$\kappa(\sigma) = \exp\left(-\frac{\sigma^3}{\sigma_0^3}\right)$ for
$\sigma \ge 0$ and $\kappa(\sigma) = 1$ else and
has $\kappa(\sigma_{\mtbox{vac}}) \equiv \kappa_{\mtbox{vac}} \ll 1$. 

In the static case, the equations of motion
for the electric potentials $\Phi^a$ and for the confinement field
$\sigma$ following from eq.~(\ref{eq:lagrangian}) are:
\begin{equation}
  \label{eq:gauss}
  \vec{\nabla}\cdot\left(\kappa(\sigma)\,\vec{\nabla}\Phi^a\right) = -
  g_s \;\rho^{a} 
\end{equation}
and
\begin{equation}
\label{eq:dsigma}
  \nabla^2 \sigma =  U'(\sigma) - {\textstyle \frac{1}{2}
  \frac{\kappa'(\sigma)}{\kappa(\sigma)^2}} 
  \left(\vec{D^3}\cdot\vec{D^3} + \vec{D^8}\cdot\vec{D^8}\right),
\end{equation}
where $\vec{D^a} = \kappa(\sigma) \vec{\nabla}\Phi^a$ denotes the
color electric displacement. The energy (neglecting quark masses) is
given by: 
\begin{eqnarray}
  \label{eq:energy}
  E &=& E_\sigma + E_g\\
  \label{eq:sigmaenergy}
  E_\sigma &=& \int \left( {\textstyle \frac{1}{2}} (\nabla\sigma)^2 +
    U(\sigma) \right) d^3r\\
  \label{eq:phienergy}
  E_g &=& \frac{1}{2} \int  \left(\vec{E^3}\cdot\vec{D^3} +
    \vec{E^8}\cdot\vec{D^8}\right) d^3r \equiv
  \int \varepsilon_g \, d^3r.
\end{eqnarray}
Confinement of color fields in our model is achieved by means of
Gauss's law in 
eq.~(\ref{eq:gauss}) and the characteristic form of the dielectric
function $\kappa(\sigma)$: A single colored quark would generate a
spherical electric field. In the vicinity of 
the quark the field is strong enough to push the confinement field
from $\sigma = \sigma_{\mtbox{vac}}$ towards smaller values and forms a cavity
in the surrounding vacuum. As $\kappa(\sigma)$ drops to zero at the
boundary of the cavity, the
electric field $\vec{E^a} = \frac{\vec{D^a}}{\kappa(\sigma)}$ diverges
and so does the electric field energy (\ref{eq:phienergy}). Note that
in this version of CDM there is no direct coupling between the
quarks and the confinement field as proposed in
\cite{Friedberg:1977eg,Schuh:1986mi}.

\section{$\bar{q}q$ Strings}
\label{sec:barqq-strings}

In contrast to configurations with net color, all white configurations
have finite energy and color fields are confined into well defined
spatial regions. Again, eq.~(\ref{eq:gauss}) enforces an electric
color field, but field lines now end on the anti--color and are
parallel to the boundary of the cavity. In this case the color
electric displacement is suppressed with the dielectric constant in
the non-perturbative vacuum. Both the electric field energy and the
confinement field energy are negligible in the outside.

\begin{figure}[htbp]
  \begin{center}
    \resizebox{0.4\textwidth}{!}{%
      \includegraphics{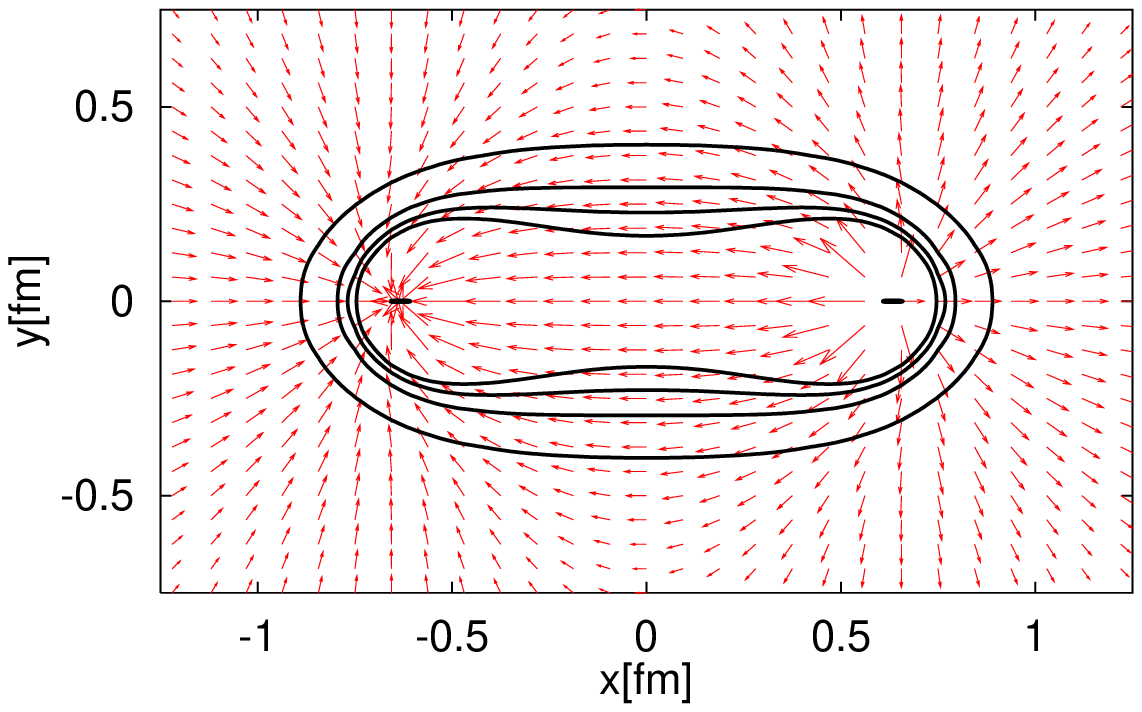}
      }
    \hfill
    \resizebox{0.4\textwidth}{!}{%
      \includegraphics{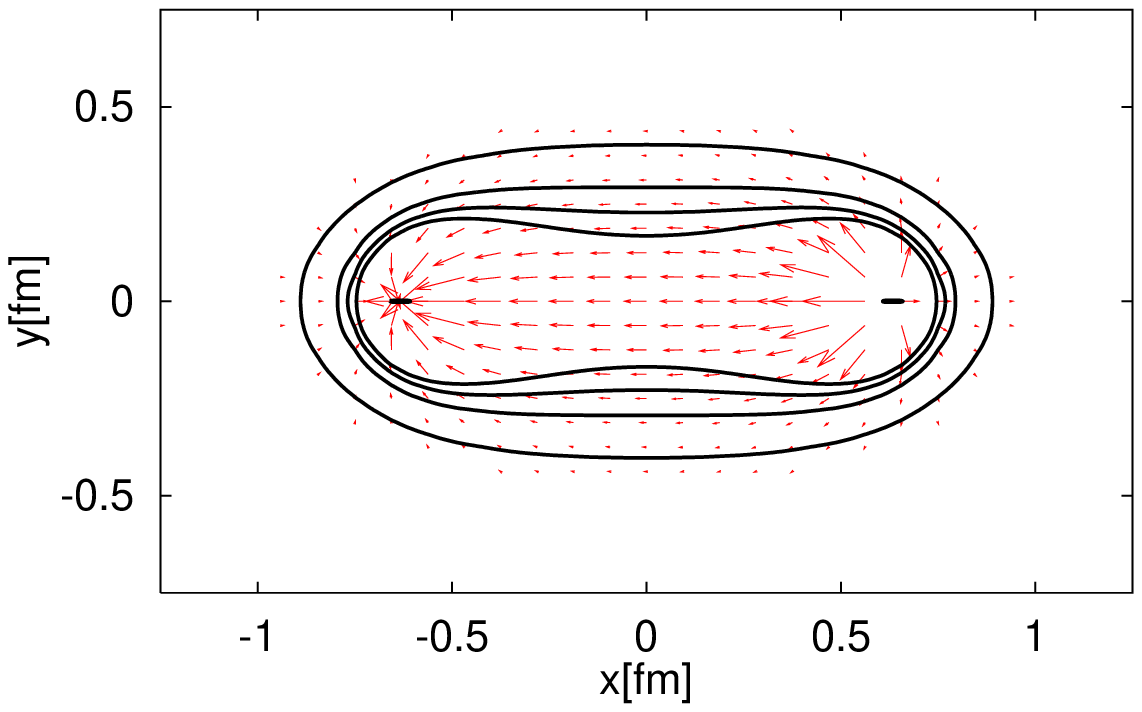}
      }    
    \caption{The color fields $\vec{E^3}$ (upper panel) and
      $\vec{D^3}$ (lower panel). The contour lines give the electric
      energy density (\ref{eq:phienergy}) $\varepsilon_g = 1,3,5,7$
      fm$^{-4}$ from the 
      outside. Only the electric displacement $\vec{D^a}$ is confined
      to the flux tube.}
    \label{fig:stringED}
  \end{center}
\end{figure}

\begin{figure}[htbp]
  \begin{center}
    \resizebox{0.24\textwidth}{!}{%
      \includegraphics{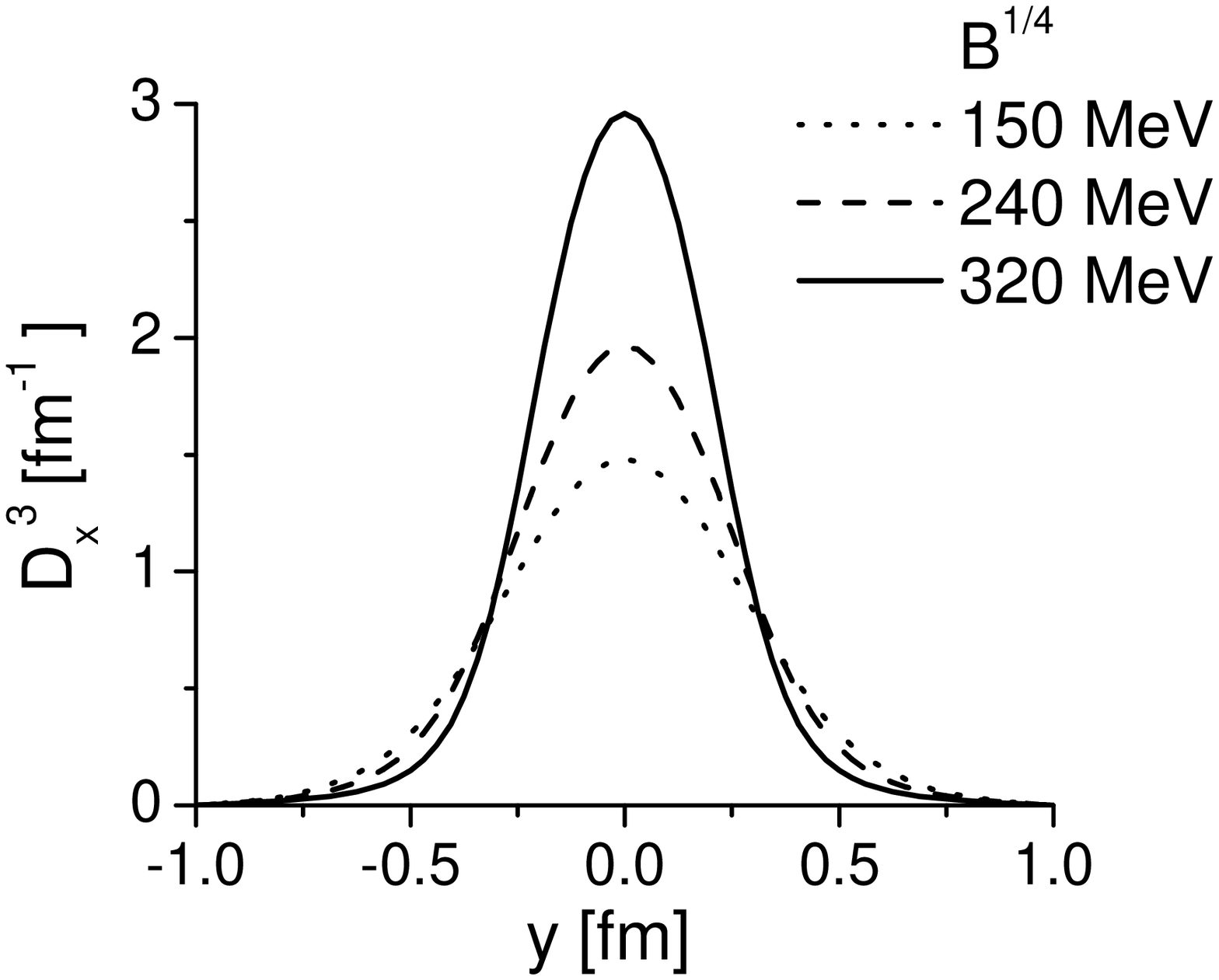}
      }
    \hfill
    \resizebox{0.24\textwidth}{!}{%
      \includegraphics{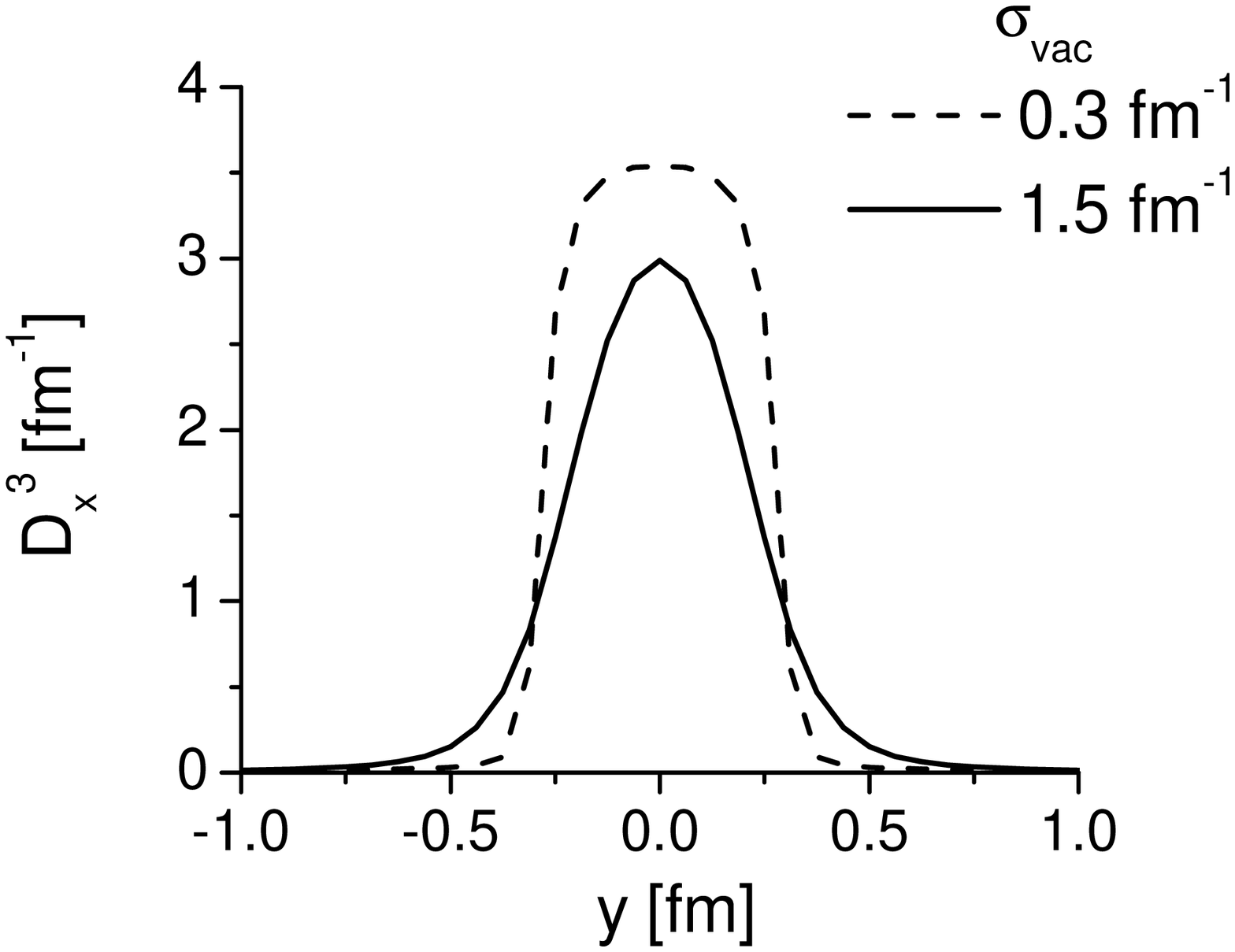}
      }    
    \caption{The string profile of a $1.2$ fm string. Increasing the
      bag pressure $B$ decreases the width and simultanously increases
      the maximal value of the $\vec{D}$ field (left) and increasing
      $\sigma_{\mtbox{vac}}$ smoothes the surface (right). All
      constant parameters are taken from tab.~\ref{tab:parameter}.
      }
    \label{fig:profile}
  \end{center}
\end{figure}

In this section we study the field configurations of color flux tubes
stretching from a quark $q$ to an antiquark $\bar{q}$. We start by
showing the electric field $\vec{E^3}$ and $\vec{D^3}$ in fig.~\ref{fig:stringED}.
It is seen that the electric displacement vanishes outside the
cavity. The  flux tube can be characterized by the profile
function, i.~e. the component of $\vec{D}$ parallel to the string axis along
the center line perpendicular to the string axis. This profile has
been studied within lQCD in \cite{Bali:1995de}.
\footnote{Note that in CDM the $\vec{D}$ field is confined and we
  compare it to the $\vec{E}$ field of reference \cite{Bali:1995de}.
}
The profile depends mainly on the choice of 
$U(\sigma)$, i.~e. on the bag constant $B$ and the vacuum
value $\sigma_{\mtbox{vac}}$ as shown in fig.~\ref{fig:profile}. The
bag constant acts as a pressure against the electric field and therefore
an increasing $B$ leads to decreasing width of the profile. In order
to fulfill Gauss's law (\ref{eq:gauss}) the  electric $\vec{D}$
field on the string axis must increase with increasing $B$. 

The value of
$\sigma_{\mtbox{vac}}$ controls the surface of the bag. Decreasing its
value leads to a sharper surface. 
In our simulations
the detailed form of the dielectric function (see
fig.~\ref{fig:ukappa}) has little effect on the profile. 

\begin{figure}[htbp]
  \begin{center}
    \resizebox{0.24\textwidth}{!}{%
      \includegraphics{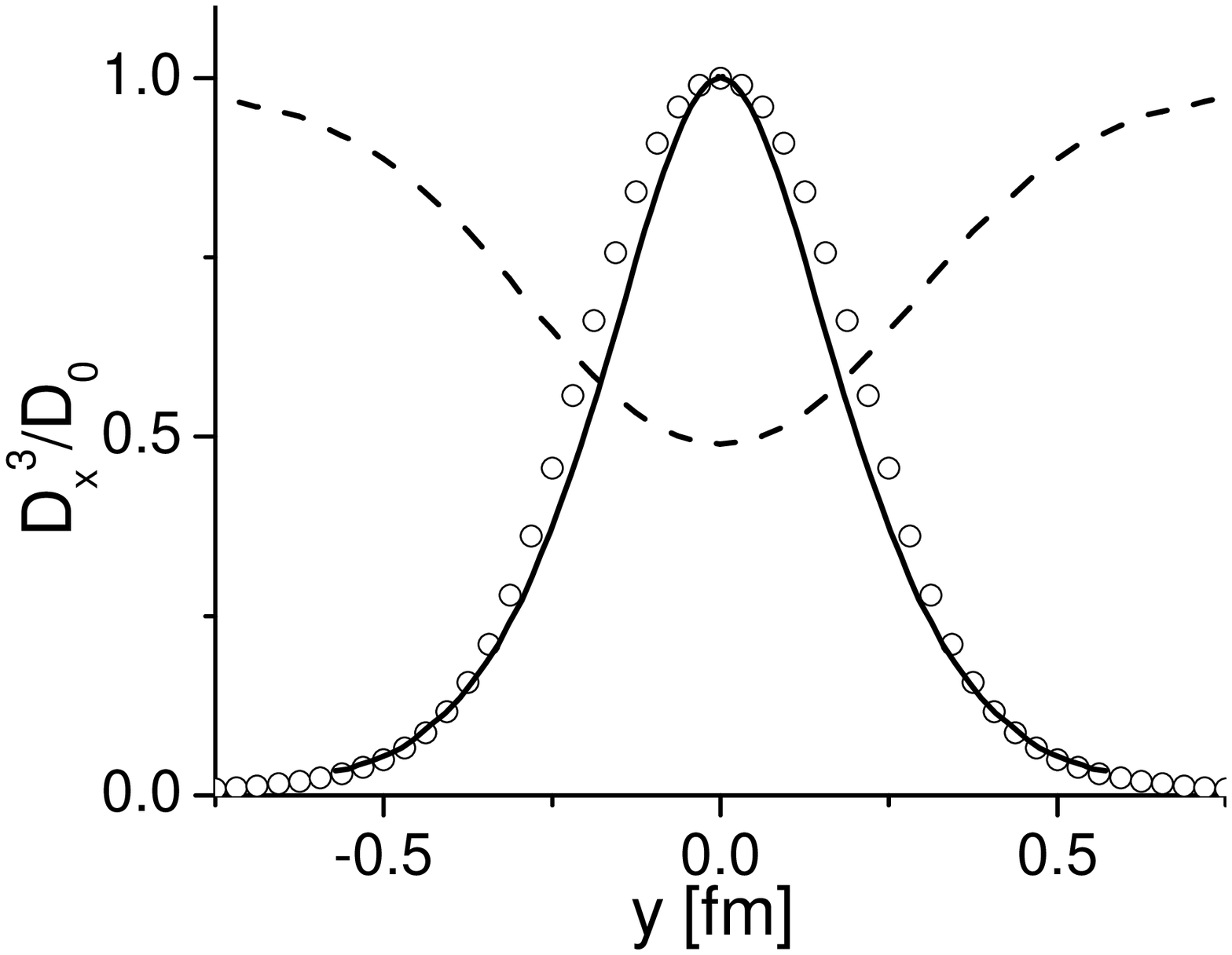}
      }    
    \resizebox{0.24\textwidth}{!}{%
      \includegraphics{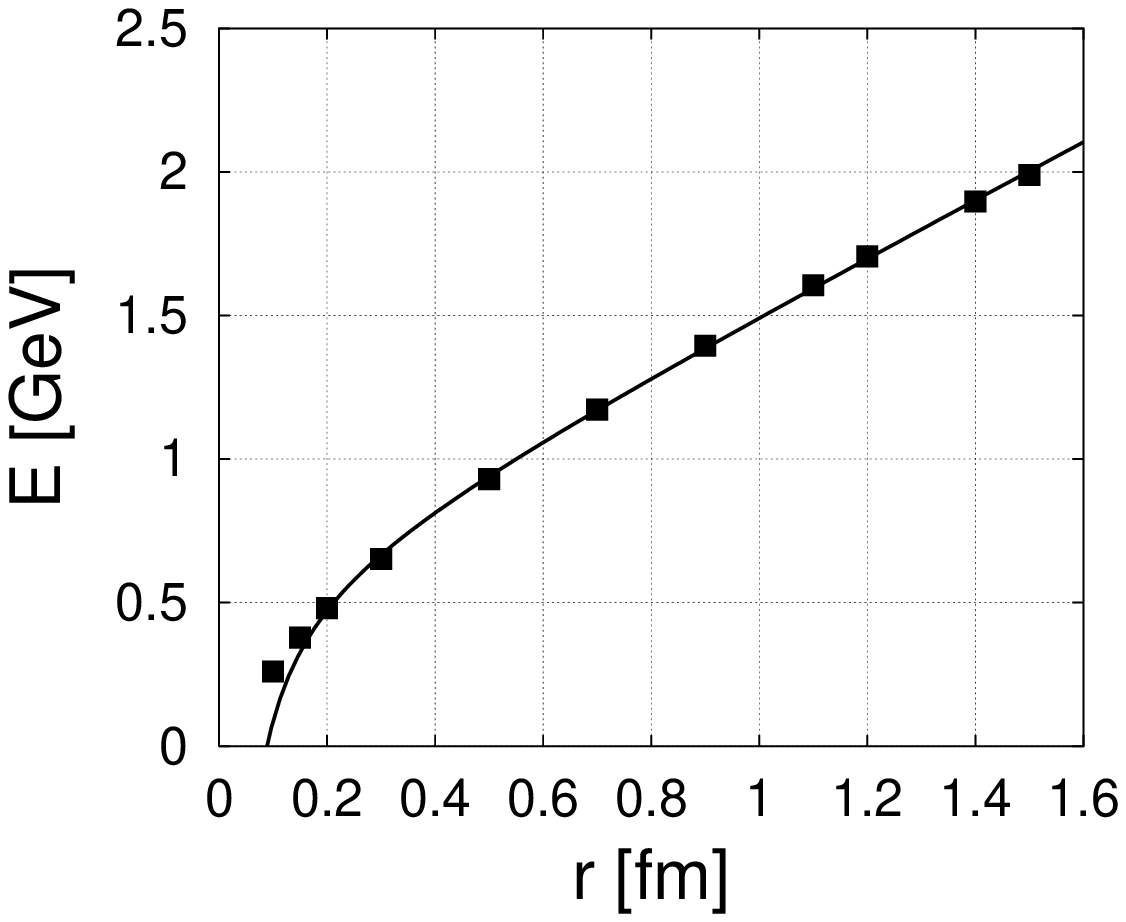}
      }        
    \caption{Left: The string profile within CDM (circles)
      compared to lQCD results (solid line)
      \cite{Bali:2000gf}. The confinement field (dashed line) drops
      to $\sigma \approx 0.5 \sigma_{\mtbox{vac}}$ inside the
      string. Right: The string potential within CDM (squares) and a
      fit to the Cornell potential (\ref{eq:cornell}). The first two
      data points are not included in the fit as the Coulomb potential
      does not hold for too small distances due to finite extension of
      the charges with rms radius $\sqrt{\langle\vec{r}^2\rangle} \approx 0.1$ fm.
      }
    \label{fig:bali_tension}
  \end{center}
\end{figure}

With the parameters given in tab.~\ref{tab:parameter} 
we reproduce the results of lQCD
\cite{Bali:2000gf,Bali:1995de} as can be seen in fig.~\ref{fig:bali_tension}.
\begin{table}[htbp]
  \begin{center}
    \begin{tabular}{cccc}
      \hline\noalign{\smallskip}
      $B^{\frac{1}{4}}$  & $\sigma_{\mtbox{vac}}$ & $\kappa_{\mtbox{vac}}$ & $g_s$  \\
      \noalign{\smallskip}\hline\noalign{\smallskip}
      $320$MeV & $1.5$fm$^{-1}$ & 0.01 & 1.0 \\
      \noalign{\smallskip}\hline
    \end{tabular}
    \caption{Model parameters used in this work.}
    \label{tab:parameter}
  \end{center}
\end{table}

Using the same parameters we can calculate the string tension of the flux
tube. We vary the $q$-$\bar{q}$ distance $r$ and plot the total energy of
eq.~(\ref{eq:energy}) as a function of $r$ in fig.~\ref{fig:bali_tension}.
For $r>0.5$ fm the energy rises linearly. We fit our results to a
Cornell potential
\begin{equation}
  \label{eq:cornell}
  E_c(r) = E_0 + \tau r - \frac{\alpha_{\mtbox{eff}}} {r},
\end{equation}
where the linear term reflects the confinement behavior for large
$\bar{q}q$-separations and the Coulomb term describes the one gluon
exchange dominant at small $r$. The constant term $E_0 = 560$ MeV is due to
electric self energies included in eq.~(\ref{eq:phienergy}).
We find a string tension $\tau = 988$ MeV/fm and a value
$\alpha_{\mtbox{eff}} = 0.291$ which is to be compared to 
lQCD results where $\alpha_{\mtbox{eff}} = 0.295$
\cite{Bali:2000gf}. 
It should be noted, that due to the high bag
pressure $B$ the electric fields are not strong enough to expel
totally the non-perturbative vacuum out of the string. The confinement
field only drops to $\sigma \approx 0.5\sigma_\mtbox{vac}$, i.~e. the
dielectric function rises to $\kappa \approx 0.5$. However,
confinement is still achieved as the energy of the color fields does
not leak into the outside.

\section{Baryons}
\label{sec:baryons}
In this section we study color fields of baryon
like $qqq$--configurations. Given that the energy scales linearly with
the $\bar{q}q$-separation, one can argue that configurations with
3 quarks sitting on the corner of an equilateral triangle will form
strings with minimal total string length. This would be a
configuration with a central \emph{Steiner} point, called a $Y$
configuration. However, if only two quark interactions are dominant,
one might expect strings stretching pairwise from one quark to
another, which would be the $\Delta$ configuration. In lQCD the
$qqq$ potential has been studied and there are indications for both
the $\Delta$ baryon \cite{Bali:2000gf,Alexandrou:2001ip} and the $Y$
baryon \cite{Takahashi:2002bw}. However in the Gaussian Stochastic
Vacuum model \cite{Kuzmenko:2000rt} clear evidence for the $Y$ ansatz
is found. 

\begin{figure}[htbp]
  \begin{center}
    \resizebox{0.24\textwidth}{!}{%
      \includegraphics{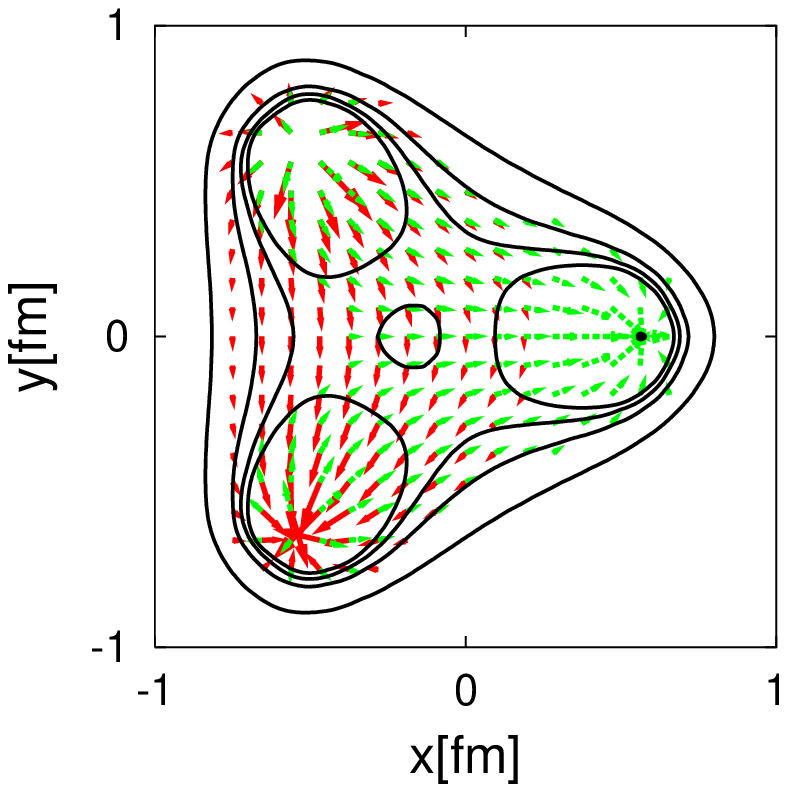}
      }    
    \resizebox{0.24\textwidth}{!}{%
      \includegraphics{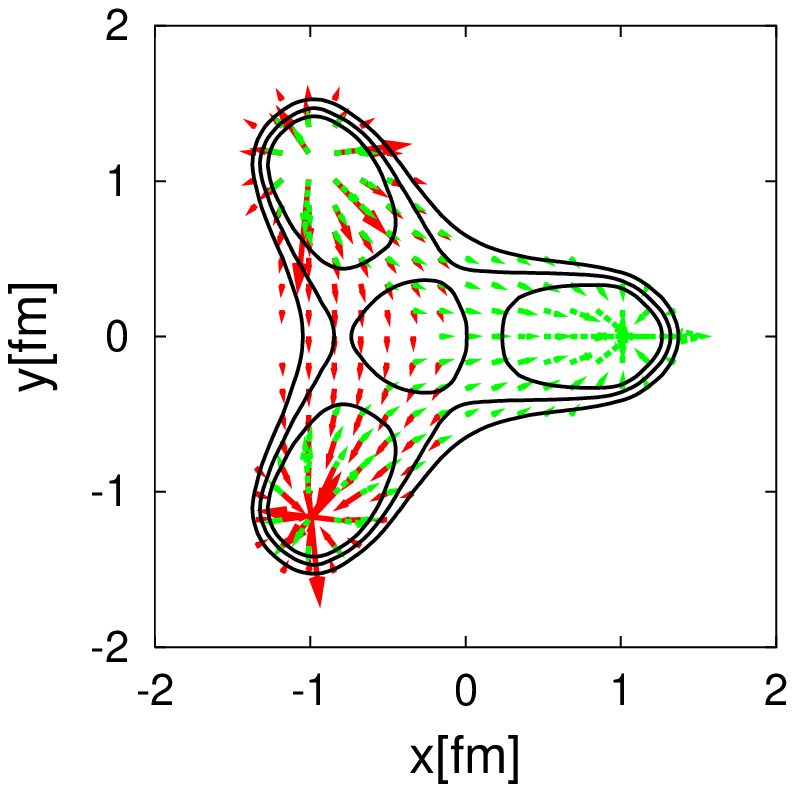}
      }    
    \caption{The color fields of a $qqq$ configuration with $\ell =0.7$($1.3$)
      fm left (right). The contour lines correspond to $\varepsilon_g = 1,3,5,7$
      fm$^{-4}$ (left) and $\varepsilon_g = 1,1.4,1.8$ fm$^{-4}$ (right).}
    \label{fig:baryon_field}
  \end{center}
\end{figure}

In fig.~\ref{fig:baryon_field} we plot the electric field distribution
for the baryon with the parameters given in
tab.~\ref{tab:parameter}. The quarks are separated a distance $\ell =
0.7 (1.3)$ fm from the Steiner point, i.e.~ the $qq$ distance is $L =
\sqrt{3} \ell \approx 1.2 (1.7)$ fm. 

The field is clearly different from a simple superposition of 3 flux
tubes between the quarks (see fig.~\ref{fig:stringED}). The electric
energy is pushed towards the center of the baryon, and a $Y$ shaped
configuration (at least for large quark separations) is seen.

\section{Summary}
\label{sec:summary}

We have analyzed the $\bar{q}q$ string within CDM and have reproduced
the geometric profile function as well as the potential. With a bag
constant $B = (320\, \mbox{MeV})^4$ and $\sigma_{\mtbox{vac}} = 1.5$ fm$^{-1}$
we get a string tension $\tau = 988$ MeV/fm and an effective strong
coupling $\alpha_{\mtbox{eff}} = 0.29$. $qqq$ configurations with
large $qq$ separations tend to show a $Y$ shaped geometry.

\end{document}